\documentclass[preprint,5p,times,twocolumn,superscriptaddress,showpacs,showkeys,preprintnumbers,nofootinbib]{elsarticle}
\usepackage{amsmath,amsfonts,amssymb,amsbsy}
\usepackage{mathrsfs,latexsym}
\usepackage{graphicx}
\usepackage{color,colortbl}
\definecolor{darkred}{rgb}{0.4,0.0,0.0}
\definecolor{darkgreen}{rgb}{0.0,0.4,0.0}
\definecolor{darkblue}{rgb}{0.0,0.0,0.4}
\usepackage[bookmarks,linktocpage,colorlinks,
    linkcolor = darkred,
    urlcolor  = darkblue,
    citecolor = darkgreen
    ]{hyperref}
\usepackage{macros_alpha}
\usepackage{dcolumn}
\usepackage{booktabs}
\newcolumntype{C}{>{$}c<{$}}
\AtBeginDocument{
\heavyrulewidth=.08em
\lightrulewidth=.05em
\cmidrulewidth=.03em
\belowrulesep=.65ex
\belowbottomsep=0pt
\aboverulesep=.4ex
\abovetopsep=0pt
\cmidrulesep=\doublerulesep
\cmidrulekern=.5em
\defaultaddspace=.5em
}
\usepackage{multirow}

\usepackage{defs}

\journal{Physics Letters B}

\begin{document}

\newcommand{\LUV}{\mu_{\rm cut}}
\newcommand{\mupt}{\mu_{\rm PT}}
\newcommand{\muhadi}[1]{\mu_{\rm had #1}}
\newcommand{\muref}{\mu_{\rm ref}}

\begin{frontmatter}

\title{{\small \phantom{.} \hfill DESY-21-177}\\[3ex]
The asymptotic approach to the continuum of lattice QCD spectral observables
}

\author{Nikolai~Husung}
\author{Peter~Marquard}
\address{Deutsches Elektronen-Synchrotron DESY, Platanenallee~6, 15738~Zeuthen, Germany}

\author{Rainer~Sommer}
\address{Deutsches Elektronen-Synchrotron DESY, Platanenallee~6, 15738~Zeuthen, Germany}
\address{Institut~f\"ur~Physik, Humboldt-Universit\"at~zu~Berlin, Newtonstr.~15, 12489~Berlin, Germany}

\begin{abstract}

We consider spectral quantities in lattice QCD and determine the asymptotic behavior of their discretization errors. Wilson fermion with 
$\rmO(a)$-improvement, (M\"obius) Domain wall fermion (DWF), and overlap Dirac operators are considered in combination with the commonly used gauge actions. Wilson fermions and DWF  with domain wall height $M_5=1+\rmO(g_0^2)$ have the same, approximate, form of the asymptotic cutoff effects: $ K\,a^2\left[\gbar^2(a^{-1})\right]^{0.760}$. A  domain wall height $M_5=1.8$, as often used, introduces large mass-dependent $K'(m)\,a^2\left[\gbar^2(a^{-1})\right]^{0.518}$ effects.
 Massless twisted mass fermions have the same form as
 Wilson fermions when the Sheikholeslami-Wohlert term \cite{Sheikholeslami:1985ij} is included. For their mass-dependent cutoff effects 
     we have information on the exponents $\Gammahat_i$
     of $\gbar^2(a^{-1})$ but not for the pre-factors. For staggered fermions there is only partial information on the exponents.  \\
We propose that tree-level $\rmO(a^2)$ improvement, which is easy to do \cite{Alford:1996nx}, should be used in the future --  both for the fermion and the gauge action. It improves the asymptotic behavior in all cases. \end{abstract}


\end{frontmatter}

\section{Introduction}
Quantum Chromo Dynamics (QCD), the fundamental theory of hadrons and nuclei can be treated perturbatively at small distances (or large momentum transfers, $\mu$), where the running coupling $\alpha_s(\mu)$ is small (asymptotic freedom). 
Discretizing the theory  on a regular space-time  lattice with spacing $a$ provides further a rigorous definition of QCD in the limit $a\to0$. 
This lattice approach has been developed in the last decades and for many observables $\obs$ a numerical ``computation'' by Monte Carlo simulations yields rather precise results (in large volume) for lattice spacings 
of $a \approx 0.1\fm \ldots 0.04\fm$. Since these 
spacings are not orders of magnitude smaller than
the typical QCD scales of order $1\fm$, care has to be taken to understand the discretisation errors 
\begin{equation} \label{eq:cutoffeffect}
  \Delta_\obs(a)  = \obs(a) - \obs(0)\,, 
\end{equation}
and remove them by an extrapolation $\obs(0)=\lim_{a\to0}\obs(a)$. 

So far these extrapolations of numerical data have used the form of
the discretisation errors of the classical theory.
Results obtained by different discretizations and 
different collaborations do not always agree at the level of 
the quoted uncertainties. For example we may look at a
presently much discussed 
 observable, the 
hadronic vacuum polarization contribution to the anomalous magnetic moment of the muon \cite{Aoyama:2020ynm,Borsanyi:2020mff}
The contribution from intermediate distances displayed in figure~4 of ref.~\cite{Borsanyi:2020mff} differs from computation to computation by much more 
than the error bars. 
More cases can be found in the FLAG report on phenomenology-related lattice results~\cite{FlavourLatticeAveragingGroup:2019iem}. Even if $\Delta_\obs$ is not the only difficult-to-control uncertainty it is certainly a very important one.

In particular, work by Balog, Niedermayer \& Weisz \cite{Balog:2009np,Balog:2009yj} in the also asymptotically free 
two-dimensional O(3) sigma model has shown that 
continuum extrapolations can be {\em very} difficult. A purely classical form of discretisation effects $\Delta_\obs(a)=k a^2 +\rmO(a^4)$ is completely off in this model. At the same time this work has realized how to use asymptotic freedom to systematically derive the leading 
asymptotic behavior. 

It should not come as a surprise that the asymptotics can be obtained analytically because
of the vanishing of the coupling for $\mu =a^{-1} \to \infty$. The steps to derive the analytic form are 1) formulate an effective field theory for the $a$-expansion. It is Symanzik's effective theory, SymEFT~\cite{Symanzik:1979ph,Symanzik:1981hc,Symanzik:1983dc,Symanzik:1983gh,Weisz:2010nr}, given by insertions of dimension six local operators (we assume $\rmO(a)$-improved QCD throughout) into the continuum path integral
2)
obtain the coefficients of these operators by matching effective theory and continuum theory at the renormalization scale $\mu=a^{-1}$, and 3) rewrite the result in terms of (in general non-perturbative) renormalization group invariants and coefficient functions which run with the scale $a^{-1}$. 

We have explained these steps in detail in \cite{Husung:2019ytz} for the example of the Yang-Mills theory. Here we define the ingredients in the final formula,~\footnote{We denote by $b_0=(11-2\nf/3)/(4\pi)^2$ the lowest order coefficient of the $\beta$ function.}
\bes \label{e:master}
   \Delta_\obs \sim a^2 \sum_{i} \chat_i  
   \left[\,2b_0 \gbar^2(a^{-1})\right]^{\Gammahat_i}  \melrgi_{\obs,i}\,,
\ees
give numerical 
results and discuss them for relevant discretizations.

Observables $\obs$ are given by (functions of) correlation functions
$\langle O \rangle$, where $O$ may be a combination of local fields and $\langle . \rangle$ denotes the Euclidean lattice path integral. 
The SymEFT  local continuum Lagrangian,
\bes
  \lag{Sym}=\lag{cont}+a^2 \sum_i \csym_i(g^2) \base_i(x)+\ldots 
\ees
describes the $a$-dependence of a scale independent $\langle O\rangle$ through local operators. 
First order 
corrections in $a^2$ are  given by the insertions,
\bes
\label{e:melements}
  \mel_{O,i}(\mu) = -\int \rmd^4 y\,\langle O \, \base_i(y;\mu)\rangle_\mathrm{cont}\,,
\ees
of the dimension six operators $\base_i$ renormalized at a scale $\mu$.
These and therefore also $\mel_{\obs,i}(\mu)$ are turned into  scale-independent ones by 
\bes
     \melrgi_{\obs,i}  = \left[\,2b_0 \gbar^2(\mu)\,\right]^{-\gammahat_i} \mel_{\obs,i}(\mu)\times[1 + \rmO(\gbar^2(\mu))]
 \ees
where $\gammahat_i$ are the leading anomalous dimensions, 
\bes
    \mu \frac{\rmd}{\rmd\mu} \base_i(y;\mu) = 2b_0\,\gbar^2(\mu)  \,\gammahat_{i} \,\base_i(y;\mu) +\rmO(\gbar^4(\mu))\,\,.
\ees
The final ingredient in \eq{e:master} is
\bes
	\Gammahat_i=\gammahat_i+n^\mathrm{I}_i\,,
\ees
where $n^\mathrm{I}_i$ give the leading perturbative behavior of
\bes 
\label{e:chat}
\csym_i(g^2)&=&\chat_i g^{2n^\mathrm{I}_i}\times(1+\rmO(g^2)).
\ees
For example in a tree-level improved theory we have $n^\mathrm{I}_i > 0$
for all $i$.

The one-loop anomalous dimensions do not depend on the renormalization scheme. We obtained them from the divergences in dimensional regularization, following the strategy described 
in \cite{Husung:2019ytz}. Details as well as analytic expressions are given in \cite{H:inprep}, results for Gradient Flow obervables \cite{Luscher:2010iy} in the pure gauge theory in \cite{H:inprep2}. 
\section{Common Lattice discretisations and Symanzik effective Lagrangian} 
We start the discussion of a few commonly used
formulations 
from the 
discretized actions,
\bes
\label{e:Slat}
  S_\mathrm{lat}&=& \Sg +\Sf\,,\; \Sf =\sum_x\, \psibar(x)\,\dop_\mathrm{lat} \,\psi(x)\,, 
\ees
of massless QCD. The gauge action,
$\Sg$, is a sum of the trace of parallel transporters around the plaquettes
of the lattice for the original Wilson action \cite{Wilson:1974}.  Other pure gauge actions are listed 
e.g. in \cite{Husung:2019ytz}, where also the $a$-expansion has been discussed in detail for the pure gauge theory. 
The classical $a$-expansion of $\Sg$ reads~\cite{Luscher:1985},
\bes
  \Sg = \int \rmd^4 x \left\{-\frac1{2g_0^2}\tr F^2 + a^2  \sum_{i=1}^2 \cclass_i \op_i  + \ldots\,\right\} \,,
\ees
where the ellipsis contains terms which can be eliminated by the 
equations of motion of the continuum theory as well as $\rmO(a^4)$ terms or total divergences.\footnote{For example the operator $\frac{1}{g_0^2} \tr (D_\mu F_{\mu\nu}\,D_\rho F_{\rho\nu})$ can be replaced by a fermionic one using
the equation of motion.
\label{foot2}}
The expansion coefficients of the operators
\bes 
\op_{1}&=&\frac{1}{g_0^2}\sum_{\mu,\nu,\rho} \tr (D_\mu F_{\nu\rho}D_\mu F_{\nu\rho})
\,,\;
\\
\op_{2}&=&\frac{1}{g_0^2}\sum_{\mu,\nu}\tr (D_\mu F_{\mu\nu}\,D_\mu F_{\mu\nu})\,
\label{eq:ops}
\ees
are denoted by $\omega_i$.
The operator $\op_{2}$ 
breaks continuum O(4) symmetry down to the lattice  group H(4). 

For a number of current large scale simulations of QCD, e.g. by MILC~\cite{Bazavov:2010ru}, CLS~\cite{Bruno:2014jqa} and KEK~\cite{Nakayama:2016atf}, Symanzik tree-level improved gauge actions are used where $\omega_1=\omega_2=0$ by construction~\cite{Luscher:1985}. Instead, 
  RBC/UKQCD~\cite{Blum:2014tka} and ETMc~\cite{Alexandrou:2021gqw} use the Iwasaki gauge action with
\bes
  \omega_1=0\,,\; \omega_2=-0.248\,, \quad  
\ees
while the Wilson plaquette action with
\bes
  \omega_1=0\,,\; \omega_2=1/12, \quad 
\ees
is hardly used any more. See also the discussion in \cite{Husung:2019ytz}.

The set $\{\op_1,\op_2\}$ not only appears in the classical expansion, but it is also the complete set of dimension six pure gauge operators in SymEFT. They are invariant 
under the symmetries of the lattice theory and exclude total divergence operators. Furthermore operators related to fermion ones
by the equations of motion$^\mathrm{\ref{foot2}}$ are dropped since they do not contribute to on-shell matrix elements \eq{e:melements} \cite{Luscher:1985}. The relevant lattice symmetries are 
Euclidean reflections, charge conjugation, H(4) and gauge transformations. The coefficients of $\op_i$ in the effective Lagrangian are $\bar \omega_i=\omega_i + \rmO(g_0^2)$. The leading $\omega_i$ are sufficient to determine the coeffcients $\chat_i$ when $n_i^\mathrm{I}=0$, see below.

We turn to $\Sf$ and  include in our discussion the most commonly used discretizations $\dop_\mathrm{lat}$ of the Dirac operator. 
These have the same space-time symmetries as $\Sg$ but depending on $\dop_\mathrm{lat}$ we have  different flavor
(vector and chiral) symmetries as listed in Tab.~\ref{t:theories}.

\begin{table}[hbpt]
  \caption{\label{t:theories}
           Fermion discretizations and symmetries of the mass-less Dirac operators. The $\mathrm{U}(1)_\mathrm{\tilde A}$ symmetry of staggered fermions \cite{Kogut:1975} is not to be mixed up with the continuum (broken) $\mathrm{U}(1)_\mathrm{A}$.
          }
  \def\arraystretch{1.1}  \begin{tabular}{llccl}
description	& $\dop_\mathrm{lat}$  & flavor symmetries 
	  \\ 
	  \hline\\[-2ex]
	 chiral &$\dop_\chi$  & $\mathrm{SU}(\nf)_\mathrm{L}\times \mathrm{SU}(\nf)_\mathrm{R}\times \mathrm{U}(1)_\mathrm{V} $ 
	 \\
	 Wilson &$\dop_\mathrm{W}$ & $\mathrm{U}(\nf)_\mathrm{V}$
	 \\
staggered &$\dop_\mathrm{st}$  & $\mathrm{U}(1)_\mathrm{B}\times \mathrm{U}(1)_\mathrm{\tilde A}$ 
	 \\
  \end{tabular}
\end{table}

``Chiral'' actions are invariant under  $\mathrm{SU}(\nf)_\mathrm{L}\times \mathrm{SU}(\nf)_\mathrm{R}\times \mathrm{U}(1)_\mathrm{V} $ for vanishing quark masses.
The symmetry transformation has a special form on the lattice~\cite{Luscher:1998pqa}, but  SymEFT is invariant under the standard  continuum transformations. All actions satisfying the Ginsparg-Wilson (GW) relation \cite{Ginsparg:1981bj} fall into this category, and Domain wall fermion discretizations (DWF) \cite{Kaplan_1992,Furman_1995,Brower:2012vk} 
can be included approximately -- usually to a very good approximation. We  need again the classical $a$-expansion
($ \sigma_{\mu\nu}=\frac i {2} [\gamma_\mu,\gamma_\nu]$)
\bes
 \dop_\mathrm{lat} &=& D_\mu \gamma_\mu - a\,\omega_\mathrm{sw} \frac{i}{4} \sigma_{\mu\nu}\,F_{\mu\nu}  \nonumber \\
 && + a^2 \omega_{3} \sum_\mu \gamma_\mu D_\mu^3 + \ldots \,,\; \quad\omega_3=1/6\,,\label{e:Dasqclass} 
\ees
of the Dirac operators given explicitly in \ref{a:dops}. 
The meaning of the ellipsis is as above.
The above simple structure is due to the form of the Wilson Dirac operator which also underlies the considered solutions of the GW relation. At this level, only the coefficient $\omega_\mathrm{sw}$ of the Pauli-term can be different. For chiral actions, it vanishes due to the GW relation and
for Wilson we assume that $\rmO(a)$ improvement has been taken care of and $\omega_\mathrm{sw}=0$. Later we will remark on the Pauli term  in connection with twisted mass fermions.
Note that $\sum_\mu \gamma_\mu D_\mu^3$ is a further source of   
O(4) violations.

Even though only one operator of dimension six appears in the classical expansion, it mixes with all other operators 
of dimension six under renormalization. Apart from the purely gluonic ones from above,
they are given by 4-fermion operators of the form
\bes
  \op_{i} &=& g_0^2 (\psibar \Sigma_i \psi)^2\,,\; 
  \\
  \Sigma_{4\ldots7}&=&
  \gamma_\mu,\,\gamma_\mu\gamma_5,\,\gamma_\mu T^a,\,\gamma_\mu\gamma_5 T^a \,,
  \\
  \Sigma_{8\ldots13}&=&
  1,\,\gamma_5,\,\sigma_{\mu\nu},\,T^a,\,\gamma_5 T^a,\,\sigma_{\mu\nu} T^a \,,
  \label{e:4ferm non-chiral}
\ees
with the standard summation convention (e.g. $\op_{13}=\sum_{\mu,\nu,a} g_0^2 (\psibar \sigma_{\mu\nu} T^a \psi)^2$).
The fields \eq{e:4ferm non-chiral} are present only for 
Wilson fermions because they are neither invariant 
under $\mathrm{SU}(\nf)_\mathrm{L}\times \mathrm{SU}(\nf)_\mathrm{R}$
nor under $\mathrm{U}(1)_\mathrm{\tilde{A}}$ transformations. Note that their tree-level coefficients vanish, 
\bes
 \omega_i=0\,,\;\;i=8\ldots 13,
\ees
for an action of the form \eq{e:Slat}.

We now turn to the effects of quark masses.
Already the continuum mass term
\bes
   \lag{\mathrm{m}}^\mathrm{cont}=\psibar m \psi \,,\quad m=\diag(m_1,\ldots,m_{\nf})
\ees
breaks the flavor symmetries down to $\mathrm{U}(1)^{\nf}$. 
The same is true for chiral and Wilson fermions and
therefore SymEFT contains all additional dimension six operators invariant under this reduced symmetry.
These eleven operators, listed in \ref{a:mops}, contain up to the third power of the mass matrix. 

For Wilson fermions, the mass-term is form-identical
to the continuum one and also the use of the  equations of motion of the massive continuum theory does not introduce any new massive operators: \bes
   \label{e:omegamassive0}
   \omega_{i}^\mathrm{W} = 0\,, \quad i\geq 14\,.
\ees
In contrast, the classical expansion of actions \eqref{e:massive_chi} with lattice chiral symmetry yields 
\bes
   \lag{\mathrm{m}}^\mathrm{\chi}&=& \lag{\mathrm{m}}^\mathrm{cont} + a^2 [\omega_{14}^\chi \op_{14} + \omega_{18}^\chi \op_{18}] +\ldots \,,
   \\
   &&\op_{14}= \frac{i}{4}\,\psibar\, m\, \sigma_{\mu\nu} F_{\mu\nu}\,\psi\,,\quad 
   \op_{18} = \psibar\, m^3\, \psi\,,
\ees
after using the continuum equations of motion.
The coefficients are 
\begin{align}
\label{e:omegamassiveGW}
   \omega_{14}^\mathrm{ov}&=1\,, &\omega_{14}^\mathrm{DWF}&=\frac{-2 (M_5-1)}{M_5(2-M_5)}
   \\
   \omega_{18}^\mathrm{ov}&=1/4\,,  &
   \omega_{18}^\mathrm{DWF}&=\frac{(M_5-1)\,(M_5^2-3M_5+1)}{M_5^2(2-M_5)^2} \,,
\end{align}
where $M_5$ is the dimensionless domain wall height for $g_0=0$, 
see \eqref{e:HDWF}.
All coefficients of massive operators vanish when $M_5$ is set to one or approaches one as $g_0^2\to0$. We include $M_5\ne 1$ because it has been used in large-scale simulations.

\section{Exponents $\Gammahat$ and coefficients $\chat$}\label{s:gammahat}

\begin{figure*} 
\begin{center} \hspace{-0.0\columnwidth}\includegraphics[width=0.99\textwidth, trim=20 0 35 15, clip]{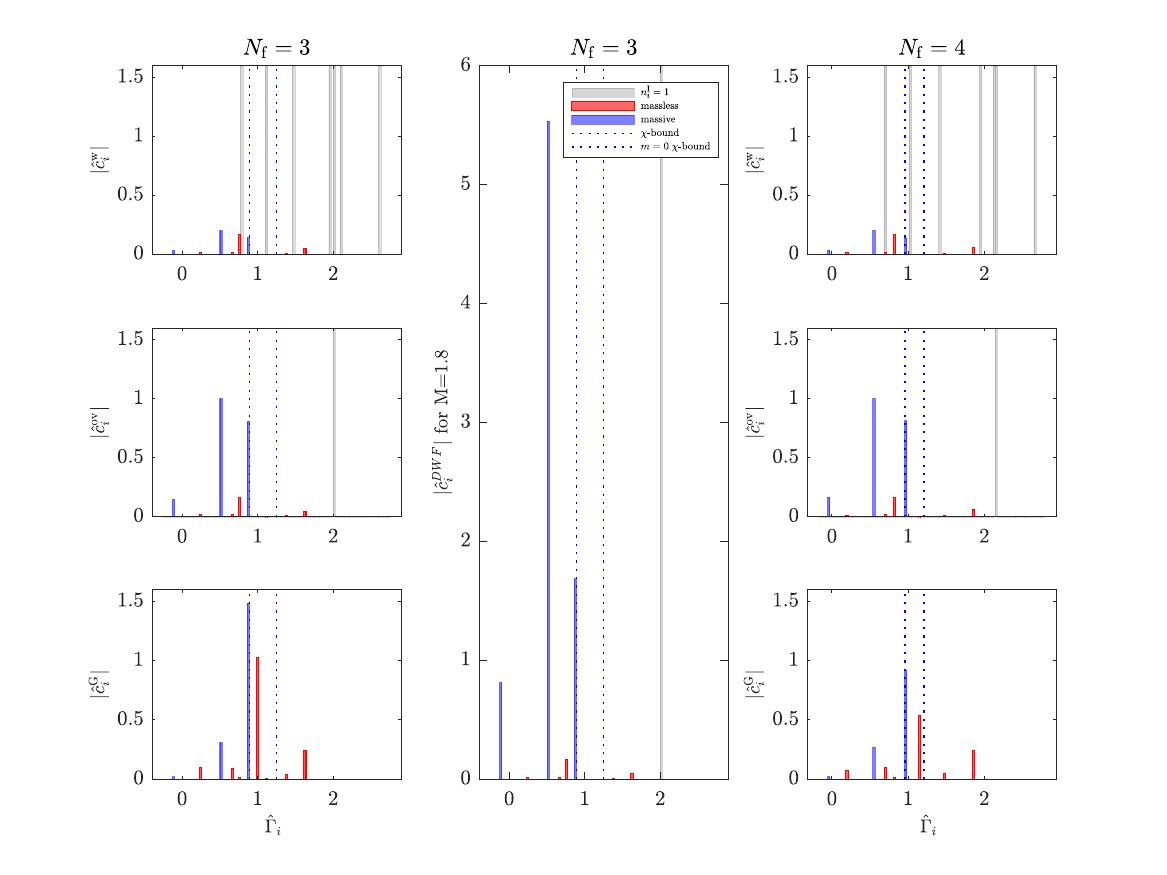}
\caption{Coefficients $\chat_i$. Gray entries just indicate the position of the unknown $\chat_i$ with $n_i=1$. \newline
\label{f:spect}}
\end{center}
\end{figure*}

The operators $\op_i$ discussed in the previous section mix under renormalization. For the central \eq{e:master} we need the coefficients of operators $\base_i$, which do not mix at one-loop 
and therefore have a unique power $\Gammahat_i$.
The renormalization of dimension six operators has been considered before, see e.g.~\cite{Jamin:1985su,Boito:2015joa,Alonso:2013hga}. Here we have to take into account a larger set including the O(4)
non-invariant operators. The computation of their one-loop anomalous 
dimensions was done along the lines of \cite{Husung:2019ytz}
and is described in detail in \cite{H:inprep}.
Here we summarize the results and provide suggestions
how to use them.

We temporarily order the operators such that the massive ones come first, 
$\opp_i=\op_{r(i)}\,,\; r(i)=i+13,\, i=1\ldots11$, 
$r(i)={i-11}\,,\; i=12\ldots24$. Then the
one-loop anomalous dimension matrix $\gammahatp$, defined by
\bes
    \mu \frac{\rmd}{\rmd\mu} \opp_i(\mu) = 2b_0\,\gbar^2(\mu)\,\sum_j\,\gammahatp_{ij}   \,\opp_j(\mu) +\rmO(\gbar^4(\mu))\,\,,
\ees
has the block structure
\begin{equation}
\eta =\begin{pmatrix}
	\gammahatp_\mathrm{mm} & 0 & 0 \\
	\gammahatp_\mathrm{\chi m} &\gammahatp_\mathrm{\chi \chi} & 0 \\
	\gammahatp_\mathrm{wm} &\gammahatp_\mathrm{w\chi} &\gammahatp_\mathrm{ww} 
\end{pmatrix} .
\end{equation}
Blocks are vanishing because massless operators do not mix into massive ones and chirally non-invariant operators do not mix into chirally invariant ones. We diagonalize $\gammahatp$ in the form
\begin{equation}
	V\,\gammahatp\, V^{-1} = \diag(\lambda_1,\ldots,\lambda_{24})\,,
\end{equation}
with a lower triangular block matrix $V$. It is composed of the left eigenvectors of $\gammahatp$ and lets us rewrite the full $d = 6$ Lagrangian at leading order in the coupling as 
\begin{equation}
	\lag{d=6}=\sum_i \omega_i\opp_i=\sum_i\csymp_i\basep_i\,,\quad \csymp_i = \sum_j \omega_{r(j)} (V^{-1})_{ji} \,.
\end{equation}
Somewhat arbitrarily we normalize
\begin{equation}
 \basep_i = V_{ij}\opp_j
\end{equation}
to the leading $\opp_j$: for each $i$ the largest absolute value of $V_{ij}$ is set to one.

In the massive theory, there are degeneracies 
$\lambda_i= \lambda_{j\ne i}$. Such terms contribute with the same power $\gbar^{2\Gammahat_i}$ in eq.~\eqref{e:master}. We thus add them up and order the eigen-values at the same time,
\bes
  \gammahat_{p(i)} &=&\lambda_i < \gammahat_{p(i)+1}.
  \\
  \base_j &=& \csym_j^{-1} \sum_{\{i|p(i)=j\}}   \bar c'_i\base'_i \,,
\ees
with the factor $\csym_j$ chosen again 
to normalize to the leading $\opp_j$.
Note that the case $\bar c'_i=0 \;\forall\;i| p(i)=j$ 
occurs only for non-degenerate eigen-values. In that case we have no contribution from $\base_j$ at leading order in $g^2$, which means that $n_j^\mathrm{I}\geq 1$ in eq.~\eqref{e:chat} and we assume
$n_j^\mathrm{I}= 1$. The coefficient $\chat_j$ is then not known. Otherwise, we have $\chat_j=\csym_j(0)$ and arrive at the ingredients $\chat_j,\Gammahat_j$ of \eq{e:master}.

Numerical results for $\nf=3,4$ are compiled in table \ref{t:results} and illustrated in figure \ref{f:spect}. Additionally we give as an example the massive $\base_i,\,i=1,3,7$ in table \ref{t:basis} for the case of Wilson quarks.
They contribute only 
when quark masses are non-zero and are marked by b=m in table \ref{t:results} and by blue entries in \fig{f:spect}. Large values are encountered for
$\chat^\mathrm{G}_i$ and the massive $\chat^\mathrm{ov}_{1,3,7}$.  Very large ones for the massive
$\chat^\mathrm{DWF}_{1,3,7}$ if $M_5=1.8$:
\bes
  \chat^\mathrm{DWF}_{1,3,7} = 
  \begin{cases}
   (-0.814,\;  -5.533,\;  -1.686 ) \;\text{ for }\; \nf=3   
   \\ 
    (-0.922,\;   -5.533,\;  -1.692	) \;\text{ for }\; \nf=4   
  \end{cases} 
  .
  \label{e:massiveM5=1.8}                            
\ees

\subsection{Renormalisation schemes and $a^2 m^n$ effects}

Apart from $\op_{14}$ and $\op_{15}$ the massive $\op_i$ are  
either of the form 
\begin{equation}
	 \op_i=\frac{f_i(m)}{2g_0^2}\tr F^2 \,,\quad i=16,17\,,
\end{equation}
or
\begin{equation}
	\op_i= \psibar\, h_i(m)\, \psi \,, \quad i\geq 18\,,
\end{equation}
with $f_i(m),h_i(m)$ functions of the quark mass matrix $m$. For continuum extrapolations 
in a massive hadronic renormalisation scheme\footnote{
In a hadronic scheme, the theory is renormalised by specifying $\nf$ ratios of hadron masses (or other dimensionfull hadronic parameters). Keeping those fixed when taking the continuum limit, the quark mass matrix is eliminated.}, such 
terms are just absorbed into the renormalisation of the 
theory and do not contribute to 
 cutoff 
effects. In particular, this is so when one extrapolates to 
the continuum at the physical point or in the chiral limit. We 
keep those terms because in practice one often uses mass-independent renormalisation schemes (see the discussion in \cite{Luscher:1996sc} at the level of $\rmO(a)$) or performs 
combined extrapolations to the physical point and the continuum limit (see e.g. \cite{Bruno:2016plf}).
Then different values of the renormalized masses enter 
in one extrapolation formula and all massive terms contribute to the $a^2$ effects unless their coefficients vanish. 

A look at table \ref{t:basis} shows that $\op_{15}$
does not contribute at all and  only 
$\base_3$ contains $\op_{14}$.

\begin{table*}[t!]
  \caption{\label{t:results}
           Numerical results for $\chat$ at one-loop ordered according to $\Gammahat_i$.  
           We label by "b=m" the massive terms where 
           all components $V_{ij}=0, \; j>11$.    The coefficients $\chat^\mathrm{W}$ or $\chat^\mathrm{ov}$ assume $\omega_1=\omega_2=0$ and $\chat^\mathrm{G}$ is
           for $\omega_i=-0.248\,\delta_{i2}$, i.e. the pure contribution of the Iwasaki gauge action. 
Entries "x" are unknown because the tree-level matching vanishes.
The tree-level coefficients at $i=8$ ($\nf=3$) and $i=10$ ($\nf=4$) vanish if $\omega_2=0$. In this case,  they contribute one power in the coupling further suppressed with unknown coefficients. These contributions are listed with $i=15$ and $i=16$ respectively.
Note that $\chat^\mathrm{\chi}_i=\chat^\mathrm{ov}_i$ for all non-massive operators and any of the chiral actions considered. For $\chi=\rm DWF$ and $M_5=1$ the coefficients of the massive operators equal the Wilson ones. For $M_5=1.8$ \cite{Blum:2014tka} we have given them in \eq{e:massiveM5=1.8} .}
           \small
  \def\arraystretch{1.0}
  \begin{tabular}{r|rrrrrr|rrrrrr}
  & \multicolumn{6}{c|}{$\nf=3$} &  \multicolumn{6}{c}{$\nf=4$} \\
  \hline &&&&&&\\[-1ex]
	$i$ & $\gammahat_i$ & $\Gammahat_i$ & b & $\chat^\mathrm{W}_i$ & $\chat^\mathrm{ov}_i$ & $\chat^\mathrm{G}_i$ 
	    & $\gammahat_i$ & $\Gammahat_i$ & b & $\chat^\mathrm{W}_i$ & $\chat^\mathrm{ov}_i$ & $\chat^\mathrm{G}_i$ 
	  \\ \hline &&&&&&\\[-2ex]
  1   &  -0.11111   &  -0.11111   &    m  &  -0.02941   &  0.14706   &  0.01969   &  -0.04000   &  -0.04000   &    m  &  -0.03333   &  0.16667   &  0.02232   \\
  2   &  0.24731   &  0.24731   &     &  -0.01593   &  -0.01593   &  0.09482   &  0.20902   &  0.20902   &     &  -0.01189   &  -0.01189   &  0.07077   \\
  3   &  0.51852   &  0.51852   &    m  &  -0.20000   &  1.00000   &  0.30905   &  0.56000   &  0.56000   &    m  &  -0.20000   &  1.00000   &  0.26784   \\
  4   &  0.66790   &  0.66790   &     &  -0.01502   &  -0.01502   &  0.08937   &  0.69814   &  0.69814   &     &  -0.01624   &  -0.01624   &  0.09665   \\
  &&&&&&&&& \multicolumn{4}{c}{\dotfill ~~ w ~~\dotfill }  \\
  5   &  0.75991   &  0.75991   &     &  0.16374   &  0.16374   &  -0.01535   &  -0.30097   &  0.69903   &     &     x       &      --     &      --     \\
  &&& \multicolumn{4}{c|}{\dotfill ~~ w ~~\dotfill }  \\
  6   &  -0.20460   &  0.79540   &     &     x       &      --     &      --     &  0.81699   &  0.81699   &     &  0.16339   &  0.16339   &  -0.01405   \\
   &&& \multicolumn{4}{c|}{\dotfill ~~ $\chi$  ~~\dotfill} 
   &&& \multicolumn{4}{c}{\dotfill ~~ $\chi$  ~~\dotfill} 
   \\[0.2ex]
  7   &  0.88889   &  0.88889   &    m  &  0.14136   &  -0.80864   &  -1.48616   &  0.96000   &  0.96000   &    m  &  0.13551   &  -0.81449   &  -0.91566   \\
  8   &  1.00000   &  1.00000   &     &     0       &     0       &     1.03015     &  0.04000   &  1.04000   &     &     x       &      --     &      --     \\
  9   &  0.11111   &  1.11111   &     &     x       &      --     &      --      &  1.13963   &  1.13963   &     &  0.00085   &  0.00085   &  -0.00505   \\
  10   &  1.12600   &  1.12600   &     &  -0.00071   &  -0.00071   &    0.00422     &  1.16000   &  1.16000   &     &     0       &     0       &     0.53568      \\
   &&& \multicolumn{4}{c|}{\dotfill ~~ massless $\chi$   ~~\dotfill}    &&& \multicolumn{4}{c}{\dotfill ~~ massless $\chi$   ~~\dotfill} \\[0.2ex]
  11   &  1.37854   &  1.37854   &     &  -0.00605   &  -0.00605   & 0.03604   &  0.41903   &  1.41903   &     &     x       &      --     &      --     \\
  12   &  0.46207   &  1.46207   &     &     x       &      --     &      --     &  1.48654   &  1.48654   &     &  0.00844   &  0.00844   &      -0.05023     \\
  13   &  1.63762   &  1.63762   &     &  -0.04642   &  -0.04642   &      -0.24365     &  1.85235   &  1.85235   &     &  -0.05666   &  -0.05666   &   -0.24313   \\
  14   &  0.94534   &  1.94534   &     &     x       &      --     &      --     &  0.94097   &  1.94097   &     &     x       &      --     &      --     \\
  15   &  1.00000   &  2.00000   &     &     x       &     x       &     x      &  1.12000   &  2.12000   &     &     x       &     --     &     --   \\
  16   &  1.11111   &  2.11111   &     &     x       &      --     &      --     &  1.16000   &  2.16000   &     &     x       &     x       &     x      \\
  17   &  1.61201   &  2.61201   &     &     x       &      --     &      --     &  1.66097   &  2.66097   &     &     x       &      --     &      --   
\end{tabular}
\end{table*}

\off{
\begin{table}[t!]
  \caption{\label{t:results_massive}
           Same as table \ref{t:results}, but only the massive terms, for a DWF action with $M_5=-1.8$ as used by \cite{RBC}. }
           \small
  \def\arraystretch{1.0}
  \begin{tabular}{r|rrrrrrr|rrrrrr}
	$i$ & \multicolumn{2}{c}{$\chat_i$} \\
  \hline \\[-1ex]
  1   &  -0.81373    &   -0.92222     \\
  3   &  -5.53333   &   -5.53333      \\
  7   &  -1.68580   &  -1.69165      \\
   \end{tabular}
\end{table}
}  
  
\subsection{Numerically dominant contributions}

The general form of lattice artifacts is complicated
because several terms with similar $\Gammahat_i$ contribute. Fortunately 
 numerically small suppression factors are present which may be taken
into account.
Indeed, a look at figure \ref{f:spect} shows that 
the coefficients $\chat_i$ differ drastically in magnitude. 
We have no reason to expect such variations also for the
unknown $ \melrgi_{\obs,i}$. Despite their somewhat arbitrary normalization we assume that they are comparable. Another significant suppression factor is given 
by the light quark masses, where we assume $am_{k}\leq am_\mathrm{strange}\lesssim1/20$, which holds for reasonable lattice spacings and for the physical strange quark mass $m_\mathrm{strange}$.~\footnote{Exceptions 
are simulations using deliberately heavy quarks as a tool \cite{DallaBrida:2019mqg}.}

Two more restrictions need to be taken into account. 
First, 
each term in \eq{e:master}  receives corrections 
of order $\gbar^{2\Gammahat_i+2}$ with unknown coefficients.  
Not accounting for the quark mass suppression, we should therefore restrict the discussion to
$\Gammahat_i < \Gammahat_1+1$. That border 
is marked by "$\chi$" in tables and figures. With  
the quark mass suppression we may further ignore $i=1$ 
and use $\Gammahat_i < \Gammahat_2+1$ as the effective boundary. It is marked by "massless $\chi$".
Second, the coefficients of the chiral symmetry violating four-fermion operators which contribute only for Wilson fermions
are unknown. For strict statements we therefore remain below $\Gammahat_6$ (border "w") for Wilson fermions. 

Let us now enter a more detailed discussion of the numbers. 
\\[1ex]
\noindent{$\mathbf{\nf=2+1}$
\\[1ex]
First we consider a tree-level improved gauge action ($\omega_1=\omega_2=0$) and $\rmO(a)$ improved {\bf Wilson fermions}. Within a precision of $10^{-2}$, 
we then have a dominance of $i=5, \;\chat^\text{W}_5= 0.16, \quad \Gammahat_5=0.76$
while  $\chat^\text{W}_{2,4}$ are a factor $1/10$ smaller
in magnitude and others are even smaller or suppressed by the quark masses. This simple structure arises because only $\omega_3$ is
non-vanishing and mixing effects are relatively small.
The latter means that also $\base_5 \approx \op_3$ is not such a
bad approximation. The first term with unknown
coefficient has $\Gammahat_6=0.795$. Since it is very close to $\Gammahat_5=0.760$, it
can  effectively be absorbed into the $i=5$ term. This leaves further corrections with 
$\Gammahat_9=1.11$ and others not far above.

Next we consider actions with {\bf exact lattice chiral symmetry}. Since $\omega_3$ is unchanged,
the situation for $m=0$ is the same, except that 
the border for corrections with unknown coefficients is pushed to $\Gammahat_{10}=1.126$ and the spectrum above is less dense.
On the other hand, the $m_\mathrm{strange}$ contributions 
to $\base_{1,3,7}$ may not be entirely negligible. For DWFs with $M_5=1$ they are just due to $\omega_3$, i.e. identical to those of Wilson fermions -- very small. For overlap fermions with the Wilson kernel, the two coefficients $\chat^\mathrm{ov}_{3,7}$ are of order one, which may make up for the mass-suppression. For 
$M_5=1.8$, as used in \cite{Blum:2014tka}, these two
coefficients are even much larger and it seems like 
mass-dependent cutoff effects are very relevant. 

Without tree-level improvement in the gauge sector,
also the contribution from
the gauge action needs to be considered,
in particular due to a rather strong mixing from 
the massive operators.  
The coefficients $\chat^\mathrm{G}$ listed in the last column have to be added to $\chat^\mathrm{ov}$ or $\chat^\mathrm{w}$ or $\chat^\mathrm{DWF}$
when the Iwasaki gauge action is used. 
For other commonly used gauge actions they are not relevant. 
\\[1ex]
}

\noindent{$\mathbf{\nf=2+1+1}$\\[1ex]
There are no large  changes in the numerical values of $\Gammahat_i,\;\chat_i$ when the charm quark is added. 
For Wilson fermions, the border where coefficients 
$\chat_i$ are known is shifted down a little. 
Below it, the coefficient of a massive operator 
is largest and it  comes with an enhancement by the charm quark mass. The dominant 
term is therefore expected to be $\chat^\text{W}_3=-0.20, \; \Gammahat_3=0.56$
and a number of terms with $\Gammahat \geq 0.699$ 
follow. The first one is generated by chirally non-invariant four-fermion operators with unknown $\chat_5$ and the following two have
relatively large coefficients $\chat_{6,7}\,\grtsim0.14$.

For actions with chiral symmetry, the statements about the massive operators made for $\nf=2+1$ remain, except that there is a large enhancement by $m_\mathrm{charm}/m_\mathrm{strange}\approx 10$.

\section{Practical consequences}
What are the  consequences of our results for recent and future large scale simulations which achieve precision results? 

First we emphasize that the knowledge gained is far from complete. 
In particular in most applications one has
matrix elements of local operators or integrated correlation functions thereof. 
The logarithmic corrections which arise from the 
$a^2$ corrections to these operators have not yet been studied. 
Similarly, we here do not discuss mixed actions and
one  has to keep in mind that there are $\rmO(a^3)$
corrections for Wilson fermions and $\rmO(a^4)$ otherwise.
Still, all these limitations are no reason to ignore what is known so far. 

For staggered fermions we unfortunately have
only limited information at present. Since the chiral invariant operators contribute, the range of eigenvalues $\gammahat_i$  
covers at least  the range given for them in \tab{t:theories}. Whether particular terms are suppressed by small $\chat_i$ or whether eigenvalues significantly outside the chiral range appear cannot be said at the moment. However, we note that MILC has been
using tree-level $\rmO(a^2)$ improved actions~\cite{Follana_2007} for a while. This suppresses all terms by one 
power of $\gbar^2(a^{-1})$. It therefore appears advisable to test a few extrapolations, e.g. 
with plain $a^2$ and $a^2\gbar^2(a^{-1})$ or even
$a^2\gbar^4(a^{-1})$ until the anomalous dimension matrix exists including taste-violating operators. 

Next, consider the DWF simulations at KEK~\cite{Nakayama:2016atf} and the improved Wilson fermion simulations of CLS~\cite{Bruno:2014jqa} for $\nf=2+1$.
Neglecting the terms proportional to the
small quark masses, 
their asymptotic behavior is given by $\Gammahat_i, \; i=2,4,5$, but since 
$\chat_2,\chat_4$ are very small it is expected
that to a good approximation $i=5$ dominates 
at small but still realistic $a$. This yields 
\bes 
\label{e:asy_wils}
   \Delta_\obs \approx & K 
   \left[\,2b_0 \gbar^2(a^{-1})\right]^{0.760}  a^2
   \times\,[1+\rmO(\gbar^2(a^{-1})^{0.351} )]\,.
\ees
For DWF, $K=-0.164 \melrgi_{\obs,5}$ is given by
a single number $\melrgi_{\obs,5}\sim \Lambda^2$ 
for each observable, while for Wilson fermions $\Gammahat_6\approx\Gammahat_5$ such that we can combine both contributions
$K= \chat_6 \melrgi_{\obs,6}-0.164 \melrgi_{\obs,5}$ from  $i=5$ and $i=6$. Recall that $\chat_6$ is an unknown one-loop coefficient which is present only when chiral symmetry is violated.

The DWF simulations of RBC/UKQCD are different both because they use the Iwasaki gauge action and 
because they have $M_5=1.8$. This yields $\chat_i=\left. \chat^\mathrm{DWF}_i\right|_{M_5=1.8}+\chat_i^\mathrm{G}$ and 
\bes 
   \Delta_\obs &\approx & 
   -0.079 \left[\,2b_0 \gbar^2(a^{-1})\right]^{0.247} a^2\melrgi_{\obs,2}
   \nonumber
   \\&&
\label{e:asy_chir}
   -0.074 \left[\,2b_0 \gbar^2(a^{-1})\right]^{0.668} a^2\melrgi_{\obs,4} 
   \\&&-0.147 \left[\,2b_0 \gbar^2(a^{-1})\right]^{0.760}  a^2\melrgi_{\obs,5}\,
   + \Delta_\obs^\mathrm{massive} \nonumber
   \\
    \Delta_\obs^\mathrm{massive} &\approx  &
    -0.794 \left[\,2b_0 \gbar^2(a^{-1})\right]^{-0.111} a^2\melrgi_{\obs,1}
    \nonumber
    \\
    && -5.22 \left[\,2b_0 \gbar^2(a^{-1})\right]^{0.519} a^2\melrgi_{\obs,3} \,.
\label{e:asy_chir_mass}
\ees
It is unclear which terms will be dominant. The $a m_\mathrm{strange}$ suppression is not large enough to compensate the large pre-factors of the massive terms. 
When continuum extrapolations are carried out at fixed renormalized masses,
the effects of $\melrgi_{\obs,1}$
are absorbed into the renormalisation of quark masses and coupling constant. In such a situation, the dominance of the
$\melrgi_{\obs,3}$ term may be 
a reasonable assumption.

$\nf=2+1+1$ simulations with twisted mass Wilson fermions at maximum twist have been carried out by 
 ETMc \cite{Alexandrou:2021gqw}. 
 In contrast to earlier $\nf=2$ twisted mass simulations, they include the Pauli-term with 
 a tad-pole improved tree-level coefficient. For
 our purpose this means that not only $\rmO(a)$ effects
 are absent (which is guaranteed in any case by automatic $\rmO(a)$ improvement at maximal twist~\cite{Frezzotti_2004,SINT_2007} ),
 but also double insertions of the Pauli term contribute only at higher order in perturbation theory. Thus,
 apart from the mass-terms, 
the asymptotic behavior is similar to RBC/UKQCD, but for $\nf=4$. It is
\bes 
   \Delta_\obs &\sim& 
   -0.059 \left[\,2b_0 \gbar^2(a^{-1})\right]^{0.209} a^2\melrgi_{\obs,2}
   \nonumber
   \\&&
   -0.080 \left[\,2b_0 \gbar^2(a^{-1})\right]^{0.698} a^2\melrgi_{\obs,4} 
   \\&&-0.150 \left[\,2b_0 \gbar^2(a^{-1})\right]^{0.817}  a^2\melrgi_{\obs,6}\,
   + \Delta_\obs^\mathrm{massive}\,. \nonumber
\ees
Here the structure of the mass-terms is not known.
The anomalous dimensions are as before, but the coefficients $\chat_{1,3,7}$ are not  known, because the full
mixing matrix has not yet been evaluated in the theory with the  
symmetry of the twisted mass term. For more details see \cite{H:inprep}.

\begin{table*}[h]
  \caption{
           Massive basis in the case of Wilson quarks, $\base_i, \;i=1,3,7$ in the form $\base_i=W_{ij}\op_{j}$ . Only non-vanishing elements are listed. Other cases and the complete basis can be reconstructed from the matrix $V$ provided in the files {\texttt{V\_Nf3.txt, V\_Nf4.txt}. }}\label{t:basis}
          \small
  \def\arraystretch{1.1}
  \begin{center}  	
  \begin{tabular}{r|rrr|rrr|c}
    & \multicolumn{3}{c|}{$\Nf=3$} &  \multicolumn{3}{c|}{$\Nf=4$} \\
j  & $W_{1j}$ &  $W_{3j}$ & $W_{7j}$  & $W_{1j}$ &  $W_{3j}$ & $W_{7j}$ & $\op_j$
	  \\ \hline &&&&&& \\[-2ex]
  14  &  0 &  0.8333 &  0 &   0 &  0.8333 &  0 & $\frac{i}{4}\bar{\psi}m\sigma_{\mu\nu}F_{\mu\nu}\psi,$
 \\ 16  &  1.0000 &  -0.1471 &  0 &   1.0000 &  -0.1667 &  0 & 
 $\tr(m^2)\frac{1}{g_0^2}\tr(F_{\mu\nu}F_{\mu\nu})$ 
 \\ 18  &  0 &  1.0000 &  1.0000 &   0 &  1.0000 &  1.0000 & 
 $\bar{\psi}m^3\psi$
 \\ 20  &  -0.8889 &  0.3529 &  0.3144 &  -0.9600 &  0.4000 &  0.3542 & 
 $\tr(m^2)\bar{\psi}m\psi$ 
   \end{tabular}
  \end{center}  	
\end{table*}

$\nf=2+1+1$ simulations with standard Wilson fermions
have their own challenges. In particular one should carefully make use of a massive renormalisation and $\rmO(a)$ 
improvement scheme \cite{Fritzsch:2018kjg}. Beyond that, our results do not 
show enhanced cutoff effects of $\rmO(a^2)$.
Coefficients $\chat^\mathrm{W}_{1,3,7}$ of the 
massive operators are small. 

For new simulations with Wilson fermions, 
or DWFs whether 2+1 or 2+1+1, we suggest to 
remove $\omega_3$ by complete $a^2$ improvement at tree-level via \cite{Alford:1996nx}
\bes 
   \dop_\mathrm{W} \to \dop_\mathrm{W} - \frac{a^2}{12}\sum_\mu
   (\nabla_\mu+\nabla^*_\mu)\nabla^*_\mu\nabla_\mu\gamma_\mu\,,
\ees
where $\nabla_\mu,\nabla^*_\mu$ are the standard gauge covariant forward and backward lattice derivatives.
Then all coefficients $\csym_i(0)$ vanish at tree level and we have $n_i^\mathrm{I}=1$ for all $i$. For Domain wall fermions
it is enough to use this improvement in the kernel operator.\footnote{It may suffice to have the sum over $\mu$ extend over $\mu=1\ldots 4$ and leave the extra dimensional couplings of the $5-d$ operator untouched.
}

\section{Conclusions}

In summary, our  most robust conclusions are:
\\
1) There is no significantly negative $\Gammahat_i$ known
so far. This is good news since strongly negative $\Gammahat_i$ slow down the approach to the continuum limit and can have drastic consequences \cite{Balog:2009np}. For staggered fermions, $\Gammahat_1<-1$, say,  cannot be excluded.
However, tree-level $\rmO(a^2)$ improved actions are often used 
which already ensures $\Gammahat_i\geq\gammahat_i+1$.
\\
2) It is advisable to include tree-level $\rmO(a^2)$ improvement also for Wilson fermion and DWF simulations 
in the future. For the latter and for twisted mass fermions,
one should search for alternatives to the Iwasaki gauge action.
\\
3) For Wilson fermions or DWF with $M_5=1+\rmO(g_0^2)$ with an improved gauge action there is a clear dominance of
the asymptotic cutoff-effects by the simple form \eq{e:asy_wils}. 

For 2+1 DWF with $M_5=1.8$ and Iwasaki gauge action the prediction
is \eqref{e:asy_chir}. It is not so obvious how to truncate it to a  form with sufficiently few parameters to be used in practice. 
We argued that the $\melrgi_{\obs,3}$ term in  \eqref{e:asy_chir_mass} dominates due to its large prefactor, but the influence of other powers of $\gbar^2$ may need to be investigated
as well. 

In general, care should be
taken with continuum extrapolations; given the difficulties 
(higher powers in $\gbar^2$ and in $a$)
a verification of the universality of the continuum limit appears more important than ever.

A number of generalizations remain to be investigated:
Gradient Flow observables~\cite{Luscher:2010iy,Ramos:2015baa},  matrix elements of electromagnetic and weak currents, 
Heavy Quark Effective Theory 
and staggered quarks. For all of these, the asymptotic
$a^2$-effects have not yet been determined. Part of the necessary preparations and perturbative computations are 
in progress, but a lot remains to be done.

}

\vskip1em
\emph{Acknowledgments: }We thank Hubert Simma, Kay Sch\"onwald and Agostino Patella for useful discussions and suggestions,
Andreas J\"uttner as well as Katsumasa Nakayama  for explanations on the used DWF actions and Stefan Schaefer for comments on the manuscript.

\section{Appendix}
\subsection{Lattice Dirac operators}
\label{a:dops}
The standard, $\Oa$-improved, Wilson Dirac operator is
\cite{Sheikholeslami:1985ij,Wilson:1974}
\bes
   \dop_\mathrm{W} = 
   \frac12(\nabla_\mu+\nabla^*_\mu)\,\gamma_\mu -\frac{a}{2} \nabla^*_\mu\nabla_\mu + a\frac{i}{4}\csw \sigma_{\mu\nu}\hat F_{\mu\nu}\,,
\ees
with forward and backward covariant derivatives $\nabla_\mu,\nabla^*_\mu$, and a discretisation of 
the field strength tensor, $\hat F_{\mu\nu}(x)=F_{\mu\nu}(x)+\rmO(a^2)$. The improvement coefficient $\csw=1+\rmO(g_0^2)$
achieves $\omega_\mathrm{sw}=\rmO(g^2)$ in  SymEFT.

We further consider Dirac operators with exact on-shell lattice chiral symmetry~\cite{Luscher:1998pqa} based on the GW relation~\cite{Ginsparg:1981bj}. The massless operators  are 
\bes 
  \label{e:Dov}
  \dop_\chi(0) = \frac{k_\chi}{a}\,[1+ \gamma_5 H_\chi (H_\chi^2)^{-1/2}]\,.
\ees 
For the original overlap fermions~\cite{Neuberger_1998}  one needs to insert
\bes
   \label{e:Hov}
   H_\mathrm{ov} = \gamma_5 (a\dop_\mathrm{W}-1)\,, \quad k_\mathrm{ov}=1, \quad \csw=0\,.
\ees
For M\"obius domain wall fermions \cite{Furman_1995,Brower:2012vk,Kaplan:1992}
the kernel operator is given by ($\csw=0$)
\bes
   \label{e:HDWF}
   H_\mathrm{DWF} = \gamma_5 
   \frac{-M_5 + a \dop_\mathrm{W}}{2-M_5 + a\dop_\mathrm{W}}\,, \quad
   k_\mathrm{DWF} = \frac{M_5(2-M_5)}{2} \,. 
\ees
up to exponentially small corrections in the extent of the fifth dimension. The dimensionless domain wall height $M_5\in (0,2)$ was taken to $M_5=1$ \cite{Nakayama:2016atf} and $M_5=1.8$ \cite{Blum:2014tka} in recent large-scale lattice simulations.

The massive Dirac operators are
\bes
   \label{e:massive_chi}
   \dop_\chi(m) = \dop_\chi(0) + m\, (1-\frac{a}{2} \dop_\chi(0) ) \,.
\ees
We note that only the tree-level value of $M_5$ enters 
our considerations. Since besides $M_5=1$ also $M_5=1.8$ 
(independent of $g_0$) has been used, we consider general $M_5$, while for overlap fermions we assume that deviations from \eqref{e:Hov} vanish as $g_0\to 0$ 
(in possible future simulations).

Violations of lattice chiral symmetry, i.e. violations of \eqref{e:Dov}, due to a finite extent of the fifth dimension of DWFs are assumed to be small compared to the discussed cutoff effects. They are beyond the scope of our approach and are hopefully monitored in an independent way.

\subsection{Massive operators of SymEFT}
\label{a:mops}
The complete massive basis of operators
$\op_{i\geq14}$ reads
\begin{align*}
\op_{14}&=\frac{i}{4}\bar{\psi}m\sigma_{\mu\nu}F_{\mu\nu}\psi,&
\op_{15}&=\tr(m)\frac{i}{4}\bar{\psi} \sigma_{\mu\nu}F_{\mu\nu}\psi,\nonumber\\
\op_{16}&=\frac{\tr(m^2)}{g_0^2}\tr(F_{\mu\nu}F_{\mu\nu}),&
\op_{17}&=\frac{\tr(m)^2}{g_0^2}\tr(F_{\mu\nu}F_{\mu\nu}),\nonumber\\
\op_{18}&=\bar{\psi}m^3\psi,&
\op_{19}&=\tr(m)\bar{\psi}m^2\psi,\nonumber\\
\op_{20}&=\tr(m^2)\bar{\psi}m\psi,&
\op_{21}&=\tr(m)^2\bar{\psi}m\psi,\nonumber\\
\op_{22}&=\tr(m^3)\bar{\psi}\psi,&
\op_{23}&=\tr(m^2)\tr(m)\bar{\psi}\psi,\nonumber\\
\op_{24}&=\tr(m)^3\bar{\psi}\psi\,.&
\end{align*}

\subsection{Additional files with results}
The  matrices $V$ for the 
cases $\nf=3,4$ are given in the included files {\tt{V\_Nf3.txt, V\_Nf4.txt}}. They also contain the eigenvalues $\gammahat_i$
with more digits than given in the text.
\\[3ex]


\end{document}